# Nanoscale control over single vortex motion in an unconventional superconductor

*Sang Yong Song[a], Chengyun Hua[b], Gábor B. Halász[b], Wonhee Ko[c], Jiaqiang Yan[b], Benjamin J. Lawrie[b], Petro Maksymovych\*[a,d]*

[a]Center for Nanophase Materials Sciences, Oak Ridge National Laboratory, Oak Ridge, TN 3781, USA
[b]Materials Science and Technology Division, Oak Ridge National Laboratory, Oak Ridge, TN 37831, USA
[c]Department of Physics and Astronomy, University of Tennessee at Knoxville, Knoxville TN 37996, USA
[d]Department of Materials Science and Engineering, Clemson University, Clemson, SC29634, USA

*Email:   pmaksym@clemson.edu

Keywords: vortex, manipulation, nanoscale, superconducting, strain

To realize braiding of vortex lines and understand the basic properties of the energy landscape for vortex motion, precise manipulation of superconducting vortices on the nanoscale is required. Here, we reveal that a localized trapping potential powerful enough to pull in the vortex line can be created with nanoscale precision on the surface of an FeSe superconductor using the tip of a scanning tunneling microscope. The mechanism of tip-induced force is traced to local modification of electronic properties and reduction of the superconducting gap, most likely due to tip-induced strain. Intriguingly, the tip-induced trapping potential is much less pronounced along the twin boundaries, dramatically reducing the vortices' degree of motion relative to the surrounding lattice. By enabling nanoscale manipulation of single vortices in Fe-based superconductors, and likely similar materials with strong strain-susceptibility of the superconducting gap, our findings provide an important step toward further development of vortex-based quantum information processing.

# 1. Introduction

For a type-II superconductor exposed to a magnetic field greater than the lower critical field, quantized magnetic fluxes penetrate the superconductor.[1] Each magnetic flux quantum breaks Cooper pairs and induces a circulating supercurrent that creates a vortex line, a cylindrical region of normal metal with a radius equivalent to the superconducting coherence length.[2, 3] Due to the Lorentz force, vortex lines repel each other and arrange into a lattice structure. These vortex lines behave like interacting elastic strings and can be partially pinned and deformed by defects.[4–7]

A growing interest in understanding and controlling superconducting vortices in recent years emerges from the potential for manipulating Majorana modes that are predicted to occur in the vortex core in topological superconductors.[8–11] Control over such Majorana zero modes is crucial for the fundamental understanding of non-Abelian statistics and in the basic study of topologically protected qubits.[12–15] The spatial manipulation of Majorana modes in vortices of topological superconductors also constitutes a proposed implementation of topological quantum information processing.[16–18]

Individual vortex manipulation is challenging in both bulk superconductors and thin films, because vortex lines are partially pinned by defects and also exhibit strong vortex-vortex repulsion. Previously, several approaches to manipulate vortices through optical[19] and resistive[20] heating, mechanical stress,[21] and magnetic force[6, 22-25] have been demonstrated. Furthermore, vortices have recently been manipulated by electronic devices,[26, 27] and it has been theoretically proposed to use domain walls to manipulate vortices.[28] However, all these techniques have limited spatial resolution, and they cannot therefore be used for the individual manipulation of densely packed vortices with strong vortex-vortex interactions. Meanwhile, in topological superconductors, a closely packed vortex lattice realizes a large density of Majorana zero modes, which is critical for scalable architectures of topological qubits. Therefore, while topological protection requires the distance between Majorana zero modes (i.e., vortices) to be larger than the coherence length[10, 29], it is otherwise desirable to keep the vortices as close as possible, which points to the need for manipulating vortices on the nanoscale.

Scanning tunneling microscopy (STM) is a natural candidate to consider for this purpose owing to its sensitivity to the vortex states, atomic spatial resolution, and a growing number of reports on the observation of Majorana zero modes through tunneling spectroscopy.[10, 11, 30, 31] STM can also famously manipulate adsorbed atoms and molecules with atomic-scale precision.[32–34] Yet, in most cases - such as nonmagnetic tips and small tunneling conductance - the tip of the tunneling microscope would not exert sufficient force onto the vortex line. Therefore, previous attempts to manipulate vortices with an STM tip required local melting of the superconducting state.[20] In addition, Kremen et al. manipulated vortices with the formation of a large and strong mechanical contact (~100 nm) on niobium (Nb) and niobium nitride (NbN) thin film using a fabricated tip of a superconducting quantum interference device (SQUID) chip.[21] Although vortex manipulation has been achieved in these cases, the precise origin of the forces, as well as the applicability of these methods to dense vortex lattices remain uncertain.

Here, we reveal a possible mechanism to effectively manipulate single vortices in an unconventional superconductor. We first demonstrate that a natural point defect near a vortex in archetypal FeSe can create a metastable potential - enabling deformation of individual vortex line by tens of nanometers, much smaller than the typical vortex-vortex distances (~140 nm at 0.13 T). Based on this observation, we determined whether vortex line deformation could be induced by creating a trapping potential with an STM tip in weak point contact with a superconducting surface. On FeSe, this approach revealed unexpectedly large vortex line deformation, enabling highly controllable displacement of individual vortices. Furthermore, we confirmed that under such weak contact conditions, the spectral weight of a large gap is suppressed but a small gap remains, for two superconducting gaps in FeSe. The local attractive potential on the vortex line is easily induced in this case, compared to previously studied BCS superconductors,[35] and ultimately enables deformation of the vortex lines even in dense vortex lattices. One possible mechanism for local suppression of one of the superconducting gaps is that local strain around the STM tip selectively reduces superconducting gaps in FeSe due to several possible pair-breaking interactions. We further applied an analytical model for vortex pinning by an external potential and revealed that the deformation of the vortex line logarithmically increases as a function of contact conductance. Moreover, the contact geometry strongly affects the strength of the trapping potential. We anticipate that this approach will work on many unconventional superconductors, particularly those with multigap structures, enabling both fundamental studies of single vortex dynamics and the susceptibility of vortex bound states to local strain, along with the eventual control of vortex-vortex interactions.

## 2. Observation of apparent dimer vortex states

First, we discuss peculiar properties of vortex lines near point defects in FeSe. The freshly cleaved surface naturally contains numerous Fe vacancies and impurities that can form a weak pinning potential (Figure S1, Supporting Information). **Figure 1**a shows a differential conductance map at -2.5 mV illustrating the vortex configuration in bulk FeSe at 0.13 T. Occasionally we observe that vortices within appear to have two cores (as highlighted by the red dotted ellipse in Figure 1a), which we refer to as "dimer-like". The separation between two components of the dimer (~15 nm) is much smaller than the average vortex-vortex distance (140 nm) on the terrace. Figure 1e shows a zero-energy conductance map of the dimer-like vortex and Figure 1f indicates a spatial variation of the vortex-bound states across core 1 and core 2 (the red dashed line in Figure 1e). Each core hosts a well-defined vortex-bound state (Figure 1f). We also found that the dimer-like vortex shown in Figure 1a changed to a vortex with only one visible core (as highlighted by the red circle in Figure 1b) after a high current scan, and one of the nearest vortices approached the position of the changed vortex (blue dotted circle in Figure 1b).

Given the rarity of the dimer vortices and the transformation between single and dimer-like configurations, we interpret dimer vortices as two nearly equivalent (metastable) configurations of a single vortex line intersecting near the surface. The metastability is underpinned by the likely presence of point impurities. During the STM tip scan, the vortex crosses over between the two states, thereby providing the dimer-like appearance. Indeed, we found some relatively larger impurities compared to individual Fe vacancies near one of the cores of the dimer-like vortex (yellow arrow in Figure S2, Supporting Information). The impurities break Cooper pairs and induce in-gap states with lower energy and higher spectral weight than in-gap states from Fe vacancies (Figure S1, Supporting Information). In addition, we found that the dimer-like vortices were affected by surrounding vortices (as shown in Figure 1b and Figure S3). A natural question arises whether the STM tip itself can create a sufficiently large and strong trapping potential, so that vortices could be manipulated at arbitrary locations on the surface.

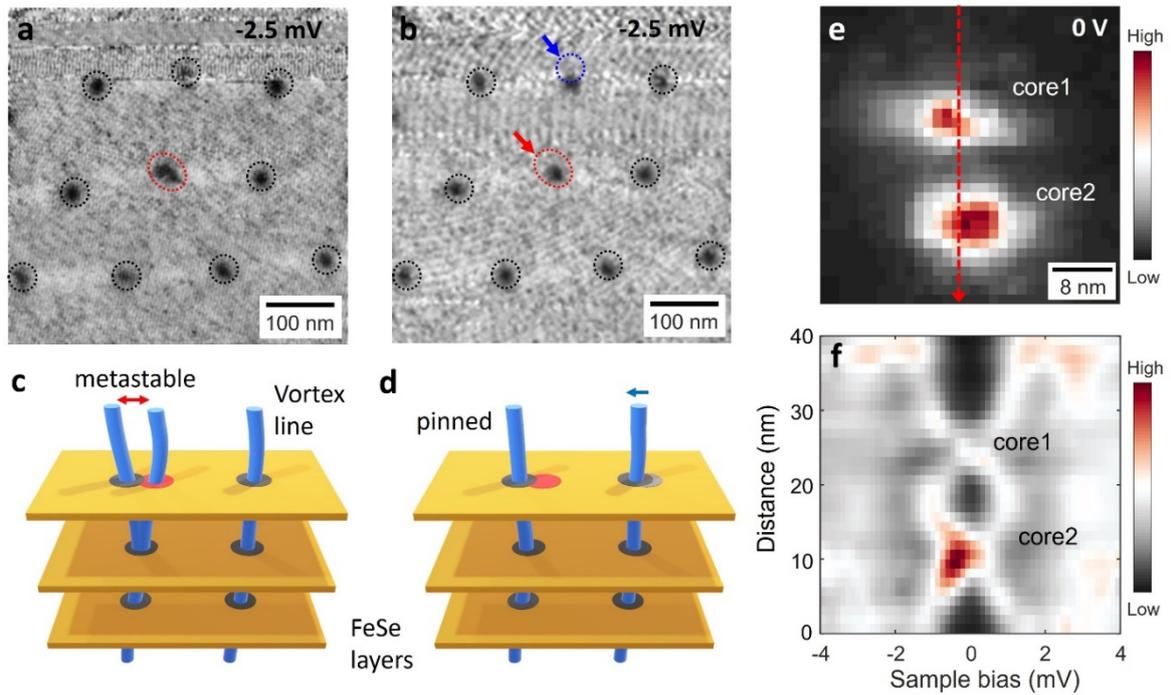

**Figure 1. Imaging of apparent dimer-like vortex on FeSe.** a) Differential conductance map of a dimer like vortex (red dotted ellipse) and surrounding vortices. b) A single vortex (red dotted ellipse) and surrounding vortices after a high current scan near the position of the dimer-like vortex. c) Schematic image of motion of a vortex line between two metastable sites (gray and red circles). d) Schematic image of the vortex line pinned at one of the metastable sites. e) Zero bias conductance map of dimer-like vortex state. f) Spatial variation of vortex bound state across the dimer like vortex (the red dashed line in (e)).

## 3. Local modification of superconducting gap

To this end, we first investigated the effect of tunneling conditions on the structure of the superconducting gap (**Figure 2**). We gradually brought the STM tip toward the surface, up to 1.2 nm from the set point $V_{bias}$ = 8 mV, and I = 1 nA (0.002 $G_0$) (z = 0 pm). Here, $G_0$ is the conductance quantum $2e^2/h$. As the STM tip approaches the surface, the conductance trace changes rapidly (blue circles in Figure 2b) and then begins to increase linearly (red circles in Figure 2b). The first abrupt change in the conductance trace (blue circles in Figure 2b) is likely due to the instability of the atomic contact condition. Moreover, the linear increase in conductance (as shown by the red circles in the inset in Figure 2b) suggests a nanoscale contact resulting from an increase in the contact area, rather than a single atomic contact. It is known that FeSe has a main gap $\Delta_1$ (~ 2 mV) and a small gap $\Delta_2$ appearing as shoulder (~ 1 mV).[36] The superconducting gaps of FeSe change significantly during tip contact to the surface. Specifically, the spectral weight of the small gap $\Delta_2$ is enhanced (blue arrows in Figure 2c) in atomic contact condition (0.13 $G_0$). Furthermore, in the regime of 0.32 ~ 1.6 $G_0$, the spectral weight of the large gap $\Delta_1$ is suppressed while the small gap $\Delta_2$ remains (red arrows in Figure 2c). We defined this regime as the strain-I regime (Figure 2b). Interestingly, when the conductance is further increased over 2.29 $G_0$, the slope of the conductance curve is changed (dark red circles in Figure 2b) and the energy position of small gap $\Delta_2$ starts to increase (dark red spectra in Figure 2c). We defined this regime (over 2.29 $G_0$) as the strain-II regime (Figure 2b). In the case of the strain-I regime, the remaining small gap $\Delta_2$ (± 1.1 mV) is reminiscent of the remaining superconducting gap at the center of a wrinkle defect (± 1.1 mV) or a twin boundary (± 1.1 mV) on the FeSe surface (Figure S4a-d, Supporting information).

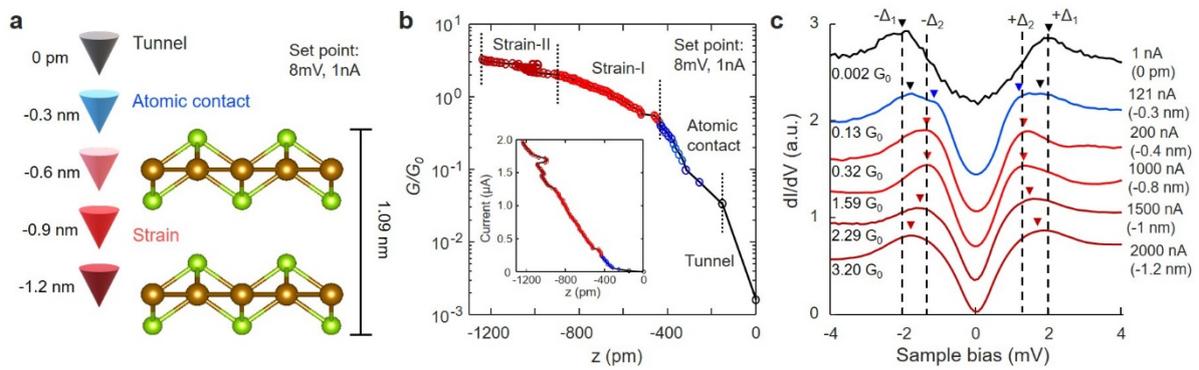

**Figure 2. STM tip induced reduction of the superconducting gap.** a) Schematic image of the tip positions relative to the FeSe atomic layers. b) Conductance and current (inset) curves with respect to the tip positions (set point: I = 100 pA, $V_{bias}$ = 8 mV). c) The variation of superconducting gaps on the FeSe surface from tunneling (0.002 $G_0$) to strain-II (3.20 $G_0$) regime. Here, the $G_0$ is the conductance quantum ($G_0 = 2e^2/h$).

### 3.1. Vortex displacement by local forces

The evolution of superconducting vortices with increasing tunneling conductance is shown in **Figure 3**. The most notable change is a dramatic enlargement of the vortex features. Figure 3a-e show differential conductance maps at -2.5 mV of the apparently expanded vortex cores as the conductance value increases from 0.51 $G_0$ (100 nA) to 3.10 $G_0$ (600 nA) (Set bias: $V_{bias}$ = -2.5 mV). The apparent size of the vortex core is reversible (Figure S5, Supporting Information) and increases from 30 nm up to as much as 140 nm at 3.10 $G_0$ on the terrace. Furthermore, we measured dI/dV spectral maps to investigate the electronic structure of the expanded vortex. **Figure 4**c,d show zero-bias differential conductance maps, which are extracted from dI/dV spectral maps of the vortex cores on the terrace and on the twin boundary (black dashed box in Figure 4a) at each starting point 0.003 $G_0$ (tunneling), and 1.57 $G_0$ (strain-I) (setpoint: $V_{bias}$ = 5 mV and 1 nA, and 600 nA, respectively). From the dI/dV spectral map, we confirmed that the apparent size of the vortex core is correlated to the magnitude of the starting conductance rather than the tunneling current at each bias (Figure S6 and S7, Supporting Information). For example, at an initial tunneling conductance of 1.57 $G_0$ (setpoint: $V_{bias}$ = 5mV and I = 600 nA), the size of the vortex is the same at 0.04 mV (0.7 nA) and at 3.6 mV (407 nA), despite two orders of magnitude difference in the tunneling current (Figure S6b,d, Supporting Information). This observation substantially rules out current-induced processes, such as current-driven force and thermal effects, as the causes of the increased apparent size of the vortex. In addition, we occasionally observed changes in the vortex array during mapping at high conductance values (Figure S8, Supporting Information).

We propose that the apparent enlargement of the vortex image corresponds to the locations where the vortex line is deformed by the attractive tip-induced trapping potential. One of the direct arguments in favor of our interpretation is that the vortex bound states are observed across the expanded vortices, as seen at the blue arrows in Figure 4h for setpoint 1.57 $G_0$ (600 nA). Figure 4g,h show the spatial variation of vortex bound states across green dotted arrows at 0.003 $G_0$ (Figure 4c) and 1.57 $G_0$ (Figure 4d), respectively. Detailed spatial variations of vortex bound states at 0.003 $G_0$ and 1.57 $G_0$ are in Figure S9 and S10. The observation of vortex bound states alone rules out spurious artifacts, such as adsorbates and topographic damage as the possible origin of the expanded vortex images

Our conclusions are further supported by the stark contrast between the vortices on the terrace and those pinned by the twin boundaries - naturally occurring planar defects due to the tetragonal to orthorhombic phase transition in FeSe below 90K.[37,38] Twin boundaries reduce the spectral weight of the large superconducting gap (Figure S4d, Supporting Information) and create a strong vortex pinning potential.[39–41] In contrast to the vortices on the terrace, vortices along the twin boundary barely expand at high conductance values in Figure 3, and Figure 4d. In addition, Figure 4f shows the spatial variation of dI/dV spectra across the vortex core on the twin boundary (orange dotted arrow in Figure 4d) at 1.57 $G_0$. On the twin boundary, the spatial distribution of vortex-bound states remains virtually unchanged in spectroscopy (black arrow in Figure 4f). The twin boundaries also reinforce the hypothesis that the tip-induced changes to the superconducting gap structure are responsible for vortex line deformations. As seen in the red spectra in Figure 4b, obtained at the center of the twin boundary (red arrow in Figure 4a), the energy position of the coherence peaks on the twin boundary are about ±1.2 mV (red arrows in Figure 4b) in the tunneling regime (0.003 $G_0$ (setpoint $V_{bias}$=5 mV, 1 nA)). The positions of the coherence peaks barely change even at high conductance conditions (red spectra in Figure 3b) unlike the coherence peaks measured on the terrace (black spectra in Figure 1b and Figure S4). Moreover, based on our earlier analysis of vortex energetics in FeSe,[41] we conservatively estimate the forces applied to the vortices to be on the order of 1-5 pN. This value is several orders of magnitude larger than the previous estimates of mechanical forces applied to isolated vortices in a conventional superconductor.[21] And we conclude that the large susceptibility of the superconducting gap in FeSe to applied strain actually enables applying the forces large enough to displace vortices within dense vortex lattices.

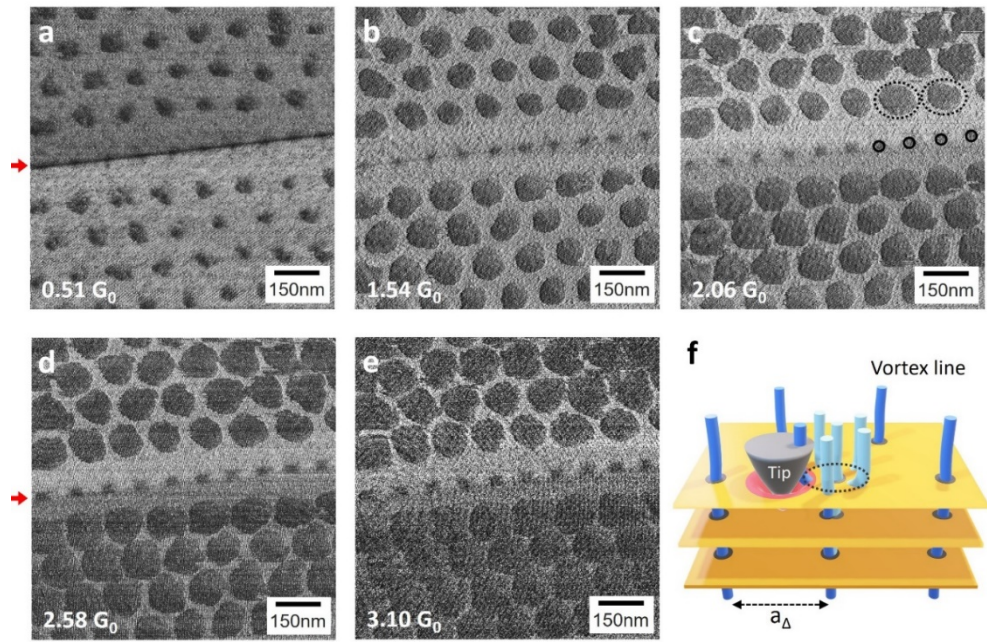

**Figure 3. Evolution of vortex images with increasing influence of the STM tip.** a-e) Large size maps of expanded vortex cores on the terrace and the twin boundary at 0.51 $G_0$ (100 nA), 1.54 $G_0$ (300 nA), 2.06 $G_0$ (400 nA), 2.58 $G_0$ (500 nA), and 3.1 $G_0$ (600 nA) measured by Tip1. f) Schematic image of deformation of vortex line near the tip induced strain field (red curved area).

We also found that the apparent size of the vortex core is sensitive to small changes of tip properties, in contrast to topographic features. We observed that a big cluster (cluster 3 in Figure S11d, Supporting Information) on the FeSe surface altered the STM tip apex. We measured differential conductance maps and topographies multiple times while the tip was altered by the cluster (Figure S11a-f, Supporting Information). Notably, the modified tip significantly affected the size of vortex cores, even at the same conductance value (Figure S11a-c, Supporting Information). However, the topographic sizes of structural clusters on the FeSe surface (cluster 1 and cluster 2 in Figure S11d-f, Supporting Information) remain constant

(Figure S11g,h, Supporting Information), regardless of the changes to the tip apex. This indicates that the tip apex is quite small, and the expanded vortex core does not originate from the tip artifact. Meanwhile, the forces on the vortices are sensitively dependent on specific tip conditions.

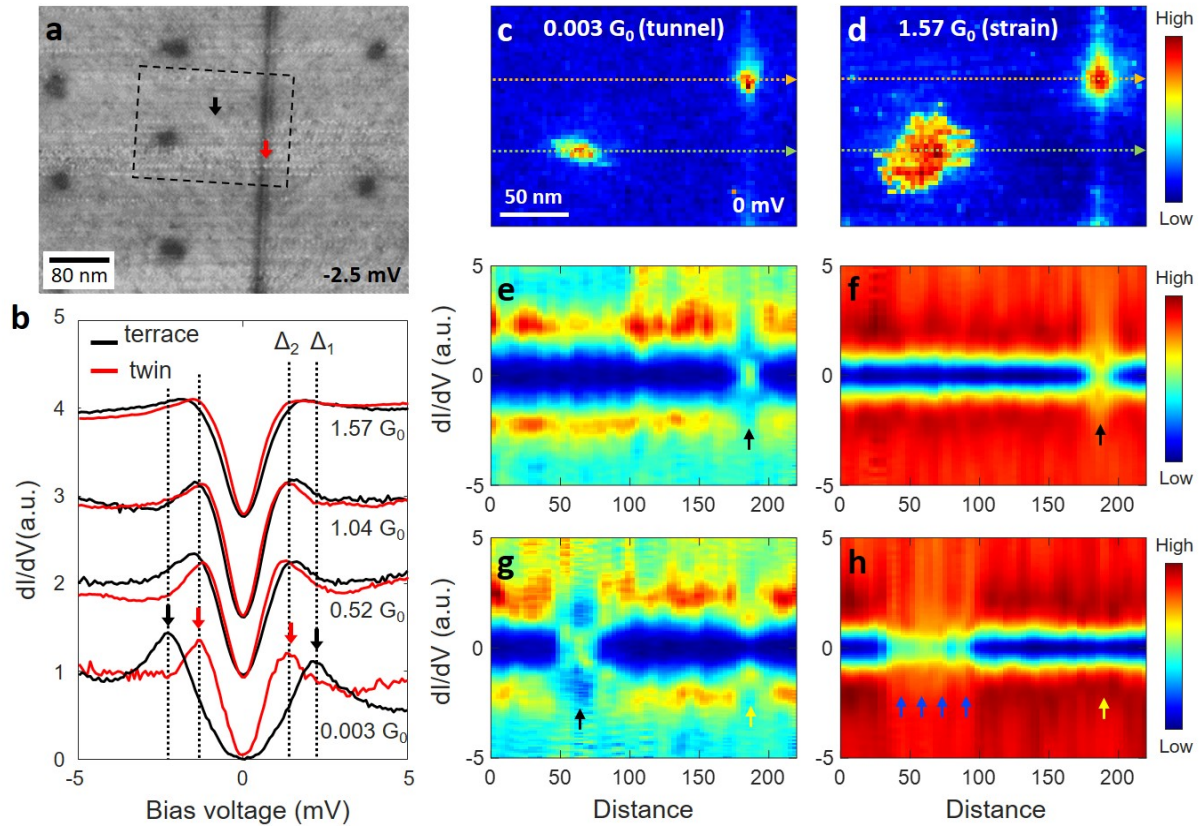

**Figure 4. Spatial variations of vortex bound states across expanded vortex cores.** a) Conductance map of vortex configuration near a twin boundary. b) Superconducting gaps at 0.003 $G_0$ (1 nA), 0.52 $G_0$ (200 nA), 1.04 $G_0$ (400 nA), and 1.57 $G_0$ (600 nA) on the terrace (black spectra) and twin boundary (red spectra). On the terrace, the superconducting gap is changed as the conductance increases, while the superconducting gap remains unchanged on the twin boundary. c, d) Zero bias conductance maps that contain a vortex on the terrace and a vortex on twin boundary at 0.003 $G_0$ (1nA), and 1.57 $G_0$ (600 nA). Here, $G_0$ is a conductance quantum $2e^2/h$. e, f) The spatial variations of superconducting gaps across the vortex on the twin boundary (orange dotted arrows in (c, d)) at 0.003 $G_0$ (1 nA), and 1.57 $G_0$ (600 nA). g, h) The spatial variations of superconducting gaps across the vortex on the terrace (green dotted arrows in (c, d)) at 0.003 $G_0$ (1 nA), and 1.57 $G_0$ (600 nA). The blue arrows indicate the core of expanded vortex.

## 3.2. Model of tip-induced trapping potential for superconducting vortices

The apparent size of the vortex image can be understood in the context of the bending of the vortex line by the trapping potential from the STM tip interacting with the surface, as schematically shown in **Figure 5**a. As defined in Ref. [42], we designate the extent of bending in the vortex line as deformation. The red, black, and blue graphs in Figure 5b show the apparent size of the vortex core on the terrace as a function of the conductance for three different STM tips. The size of the vortex core logarithmically increases with conductance, but the slope is different for different tips. We plotted graphs for the deformation (u) of the vortex line with respect to the conductance obtained from different tips (red, black, and blue circles in Figure 5c).

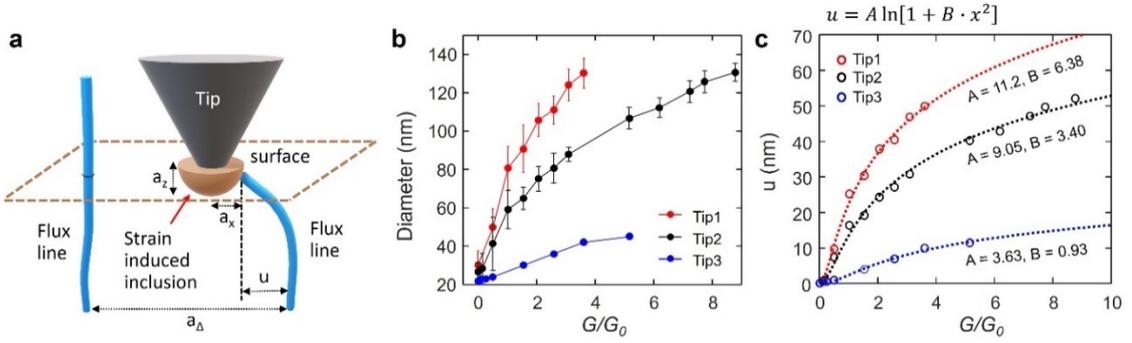

**Figure 5. Modeling STM junction as a tunable trapping potential.** a) Schematic image of tip induced inclusion and deformed vortex line. b) Dependences of the diameters of expanded vortex cores on conductance for different tip1 (red), tip2 (black), and tip3 (blue). c) Dependences of the deformations of vortex lines on the conductance values for tip1 (red circles), tip2 (black circles), and tip3 (blue circles). Each dotted line represents fits based on $A \ln[1 + B \cdot x^2]$.

We further consider the trapping potential created by the STM tip to be equivalent to an inclusion in the crystal lattice that suppresses superconductivity. In this case, the problem becomes equivalent to vortices interacting with a pinning center in the form of a non-superconducting inclusion, as previously described by Willa, R. et al..[42] The deformation of the vortex line is determined by the elasticity of the vortex line due to the vortex-vortex interaction and the maximum trapping force from the inclusion.[7, 42, 43] For inclusion sizes comparable to the coherence length ($a_x \sim \xi$), the vortex binding energy to an anisotropic inclusion is given by[43–45]

$$U_p \approx \varepsilon_0 a_z \ln\left(1 + \frac{a_x^2}{\xi_{ab}^2}\right) \tag{1}$$

Where $\varepsilon_0 = \Phi_0^2/(4\pi\lambda)^2$ is the typical vortex energy scale, $\lambda$ is the London penetration depth, $a_x$ is the inclusion size along the x-axis, $a_z$ is the inclusion size along the z-axis, and $\xi_{ab}$ is the coherence length for the ab-plane. As detailed in Supplementary note 1, we can evaluate the deformation of the vortex line using this formalism. The functional form of the tip-induced force as $A \cdot \ln(1 + B \cdot x^2)$ provides a very good fit to the experimental data. (red, black, and blue dotted lines in Figure 5c and Figure S12, Supporting Information). Moreover, in this model, we can account for the variability of the deformation by different tips, via the effective aspect ratio ($a_z/a_x$) of the tip-induced potential. When $a_x$ increases relative to $a_z$ and $a_x > a_z$, the strength of deformation decreases because of the reduced critical angle and distance at which the vortex line enters and exits the inclusion.[42] This means that the shape of the tip apex strongly affects the trapping force and the deformation of the vortex line as the tip approached. Future work will be aimed at detailed modeling of the tip-induced strain potential and its connection to the suppression of the superconductivity observed experimentally.

## 4. Summary and conclusions

To summarize, our findings show that we can deform the individual vortex lines in the surface region of FeSe with nanoscale precision using the STM tip. We revealed how point impurities and the intermediate size of the tip-induced inclusion affect the vortex line segments near the surface in FeSe. Point impurities close to the vortex core can form trapping potentials and lead to locally metastable vortex configurations, which appear as apparent dimers. Furthermore, the locally induced strain by the STM tip facilitates the selective suppression of the large superconducting gap and the generation of an effective trapping potential (inclusion) under the STM tip at arbitrary locations on the FeSe surface. Based on a framework provided by established models for pinning, anisotropy, and elasticity caused by defects, we concluded that the STM tip effectively creates a tunable, intermediate-sized anisotropic defect ($a_x \sim \xi$) from the point defect ($a_x < \xi$) as the conductance increases. We also confirmed that the strength of deformation varies significantly depending on the shape of the tip apex. Our governing hypothesis, supported by the direct spectroscopy measurement, points to local gap reduction as the driving force for the trapping potential. A method of manipulating vortices using the manipulation wrinkles, which are strained areas ($a_x \gg \xi$), has been reported[46] and strengthens our results. In FeSe thin films with weaker vortex-line energies, our method may allow for manipulation of not only the end of the vortex line but also the entire vortex line. As a result, it may be possible to implement a popular vortex braiding protocol by moving one vortex around other vortices with designed patterns[15, 47, 48]. In addition, it has been proposed that effective Majorana braiding can be achieved through much smaller displacements, very much comparable to our experiments, by moving a vortex along a small triangle defined by adjacent vortices in finite time.[49] Our results, therefore, provide additional insight into vortex pinning, vortex-vortex interactions, and vortex motion in dense vortex lattices, toward further understanding the effects of vortex-vortex interactions on accessible manipulation trajectories and the role that intrinsic vortex dynamics can play in quantum information science.

## 5. Experimental methods

We have conducted the experiment using a SPECS Joule-Thompson STM at 1.2 K in the base pressure of < $10^{-10}$ mbar. To precisely approach the STM tip, a SPECS-Nanonis controller was used. Freshly cleaved surfaces were obtained by delaminating a small sample of FeSe in low vacuum of < $10^{-7}$ mbar, followed by rapid sample transfer to the cryogenic chamber. The differential conductance maps of vortices are obtained using a lock-in technique with the modulation frequency 785 Hz.

**Supporting Information**

Supporting Information is available from the Wiley Online Library or from the author.

**Acknowledgements**

This work was supported by the U.S. Department of Energy, Office of Science, Basic Energy Sciences, Materials Sciences and Engineering Division. Scanning tunneling microscopy was performed at the Center for Nanophase Materials Sciences (CNMS), which is a US Department of Energy, Office of Science User Facility at Oak Ridge National Laboratory.

References

[1]  A. A. Abrikosov, *Zh. Eksp. Teor.* **1957**, 1442.
[2]  P. G. De Gennes, J. Matricon, *Rev. Mod. Phys.* **1964**, *36*, 45.
[3]  P. H. Borcherds, C. E. Gough, W. F. Vinen, A. C. Warren, *Philosoph. Mag.* **1964**, *10*, 349.
[4]  G. Blatter, M. V. Feigel'man, V. B. Geshkenbein, A. I. Larkin, V. M. Vinokur, *Rev. Mod. Phys.* **1994**, *66*, 1125.
[5]  G. Blatter, V. B. Geshkenbein, J. A. G. Koopmann, *Phys. Rev. Lett.* **2004**, *92*, 067009.
[6]  O. M. Auslaender, L. Luan, E. W. J. Straver, J. E. Hoffman, N. C. Koshnick, E. Zeldov, D. A. Bonn, R. Liang, W. N. Hardy, K. A. Moler, *Nat. Phys.* **2009**, *5*, 35.
[7]  W.-K. Kwok, U. Welp, A. Glatz, A. E. Koshelev, K. J. Kihlstrom, G. W. Crabtree, *Rep. Prog. Phys.* **2016**, *79*, 116501.
[8]  C. W. J. Beenakker, *Annu. Rev. Condens. Matter Phys.* **2013**, *4*, 113.
[9]  D. Wang, L. Kong, P. Fan, H. Chen, S. Zhu, W. Liu, L. Cao, Y. Sun, S. Du, J. Schneeloch, R. Zhong, G. Gu, L. Fu, H. Ding, H.-J. Gao, *Science* **2018**, *362*, 333.
[10]  T. Machida, Y. Sun, S. Pyon, S. Takeda, Y. Kohsaka, T. Hanaguri, T. Sasagawa, T. Tamegai, *Nat. Mater.* **2019**, *18*, 811.

[11]	M. Li, G. Li, L. Cao, X. Zhou, X. Wang, C. Jin, C.-K. Chiu, S. J. Pennycook, Z. Wang, H.-J. Gao, *Nature* **2022**, *606*, 890.
[12]	A. Yu. Kitaev, *Ann. Phys.* **2003**, *303*, 2.
[13]	S. Das Sarma, M. Freedman, C. Nayak, *Phys. Rev. Lett.* **2005**, *94*, 166802.
[14]	J. C. Y. Teo, C. L. Kane, *Phys. Rev. Lett.* **2010**, *104*, 046401.
[15]	X. Ma, C. J. O. Reichhardt, C. Reichhardt, *Phys. Rev. B* **2020**, *101*, 024514.
[16]	D. A. Ivanov, *Phys. Rev. Lett.* **2001**, *86*, 268.
[17]	M. Sato, S. Fujimoto, *Phys. Rev. B* **2009**, *79*, 094504.
[18]	X. Ma, C. J. O. Reichhardt, C. Reichhardt, *Phys. Rev. B* **2018**, *97*, 214521.
[19]	I. S. Veshchunov, W. Magrini, S. V. Mironov, A. G. Godin, J.-B. Trebbia, A. I. Buzdin, P. Tamarat, B. Lounis, *Nat. Commun.* **2016**, *7*, 12801.
[20]	J.-Y. Ge, V. N. Gladilin, J. Tempere, C. Xue, J. T. Devreese, J. Van de Vondel, Y. Zhou, V. V. Moshchalkov, *Nat. Commun.* **2016**, *7*, 13880.
[21]	A. Kremen, S. Wissberg, N. Haham, E. Persky, Y. Frenkel, B. Kalisky, *Nano Lett.* **2016**, *16*, 1626.
[22]	J. T. Zhang, J. Kim, M. Huefner, C. Ye, S. Kim, P. C. Canfield, R. Prozorov, O. M. Auslaender, J. E. Hoffman, *Phys. Rev. B* **2015**, *92*, 134509.
[23]	N. Shapira, Y. Lamhot, O. Shpielberg, Y. Kafri, B. J. Ramshaw, D. A. Bonn, R. Liang, W. N. Hardy, O. M. Auslaender, *Phys. Rev. B* **2015**, *92*, 100501.
[24]	A. Yagil, Y. Lamhot, A. Almoalem, S. Kasahara, T. Watashige, T. Shibauchi, Y. Matsuda, O. M. Auslaender, *Phys. Rev. B* **2016**, *94*, 064510.
[25]	E. W. J. Straver, J. E. Hoffman, O. M. Auslaender, D. Rugar, K. A. Moler, *Appl. Phys. Lett.* **2008**, *93*, 172514.
[26]	I. Keren, A. Gutfreund, A. Noah, N. Fridman, A. Di Bernardo, H. Steinberg, Y. Anahory, *Nano Lett.* **2023**, 23, 4669.
[27]	T. Golod, L. Morlet-Decarnin, V. M. Krasnov, *Nat. Commun.* **2023**, 14, 4926.
[28]	S. A. Díaz, J. Nothhelfer, K. Hals, K. Everschor-Sitte, *Phys. Rev. B* **2024**, 109, L201110.
[29]	C.-K. Chiu, T. Machida, Y. Huang, T. Hanaguri, F.-C. Zhang, *Sci. Adv.* **2020**, *6*, eaay0443.
[30]	W. Liu, L. Cao, S. Zhu, L. Kong, G. Wang, M. Papaj, P. Zhang, Y.-B. Liu, H. Chen, G. Li, F. Yang, T. Kondo, S. Du, G.-H. Cao, S. Shin, L. Fu, Z. Yin, H.-J. Gao, H. Ding, *Nat. Commun.* **2020**, *11*, 5688.
[31]	T. Zhang, W. Bao, C. Chen, D. Li, Z. Lu, Y. Hu, W. Yang, D. Zhao, Y. Yan, X. Dong, Q.-H. Wang, T. Zhang, D. Feng, *Phys. Rev. Lett.* **2021**, *126*, 127001.
[32]	J. A. Stroscio, D. M. Eigler, *Science* **1991**, *254*, 1319.
[33]	M. F. Crommie, C. P. Lutz, D. M. Eigler, *Science* **1993**, *262*, 218.
[34]	P. Maksymovych, in *Scanning Probe Microscopy of Functional Materials: Nanoscale Imaging and Spectroscopy* (Eds.: S. V. Kalinin, A. Gruverman), Springer, New York, NY, **2011**, pp. 3–37.
[35]	G. Pristáš, S. Gabáni, E. Gažo, V. Komanický, M. Orendáč, H. You, *Thin Solid Films* **2014**, *556*, 470.
[36]	P. O. Sprau, A. Kostin, A. Kreisel, A. E. Böhmer, V. Taufour, P. C. Canfield, S. Mukherjee, P. J. Hirschfeld, B. M. Andersen, J. S. Davis, *science* **2017**, 357, 75.
[37]	A. E. Böhmer, F. Hardy, F. Eilers, D. Ernst, P. Adelmann, P. Schweiss, T. Wolf, C. Meingast, *Phys. Rev. B* **2013**, *87*, 180505.
[38]	A. E. Böhmer, T. Arai, F. Hardy, T. Hattori, T. Iye, T. Wolf, H. v. Löhneysen, K. Ishida, C. Meingast, *Phys. Rev. Lett.* **2015**, *114*, 027001.

[39]  C.-L. Song, Y.-L. Wang, Y.-P. Jiang, L. Wang, K. He, X. Chen, J. E. Hoffman, X.-C. Ma, Q.-K. Xue, *Phys. Rev. Lett.* **2012**, *109*, 137004.
[40]  T. Watashige, Y. Tsutsumi, T. Hanaguri, Y. Kohsaka, S. Kasahara, A. Furusaki, M. Sigrist, C. Meingast, T. Wolf, H. v. Löhneysen, T. Shibauchi, Y. Matsuda, *Phys. Rev. X* **2015**, *5*, 031022.
[41]  S. Y. Song, C. Hua, L. Bell, W. Ko, H. Fangohr, J. Yan, G. B. Halász, E. F. Dumitrescu, B. J. Lawrie, P. Maksymovych, *Nano Lett.* **2023**, *23*, 2822.
[42]  R. Willa, A. E. Koshelev, I. A. Sadovskyy, A. Glatz, *Phys. Rev. B* **2018**, *98*, 054517.
[43]  A. E. Koshelev, A. B. Kolton, *Phys. Rev. B* **2011**, *84*, 104528.
[44]  C. J. Van Der Beek, M. Konczykowski, A. Abal'oshev, I. Abal'osheva, P. Gierlowski, S. J. Lewandowski, M. V. Indenbom, S. Barbanera, *Phys. Rev. B* **2002**, *66*, 024523.
[45]  C. J. van der Beek, M. Konczykowski, R. Prozorov, *Supercond. Sci. Technol.* **2012**, *25*, 084010.
[46]  P. Fan, H. Chen, X. Zhou, L. Cao, G. Li, M. Li, G. Qian, Y. Xing, C. Shen, X. Wang, C. Jin, G. Gu, H. Ding, H.-J. Gao, *Nano Lett.* **2023**, *23*, 4541.
[47]  C. Beenakker, *Annu. Rev. Condens. Matter Phys*. **2013**, 4, 113.
[48]  X. Ma, C. J. O. Reichhardt, C. Reichhardt, *Phys. Rev. B* **2018**, 97, 214521.
[49]  T. Posske, C.-K. Chiu, M. Thorwart, *Phys. Rev. Res*. **2020**, 2, 023205.

# Supporting Information

# Nanoscale control over single vortex motion in an unconventional superconductor

*Sang Yong Song, Chengyun Hua, Gábor B. Halász, Wonhee Ko, Jiaqiang Yan, Benjamin J. Lawrie, Petro Maksymovych\**

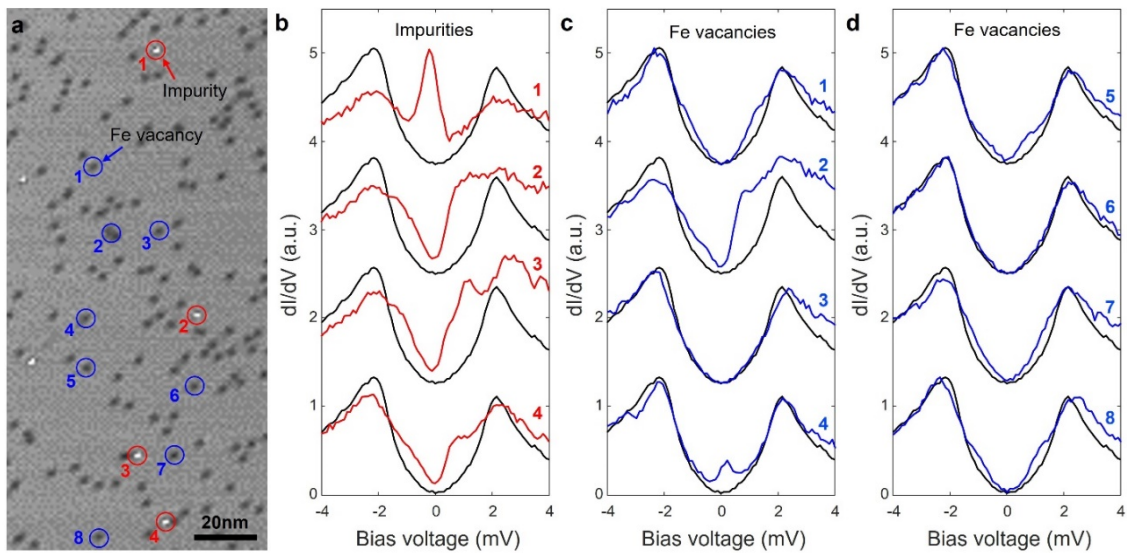

**Figure S1. Impurity bound states induced by Fe vacancies and other impurities.** a) Topographic image of Fe vacancies (black dots) and unknown impurities (white dots) at 100 mV (I = 100 pA). b) Impurity bound state inside the superconducting gap on unknown impurities. c, d) Impurity bound state on Fe vacancies. Black spectra are obtained on FeSe surface.

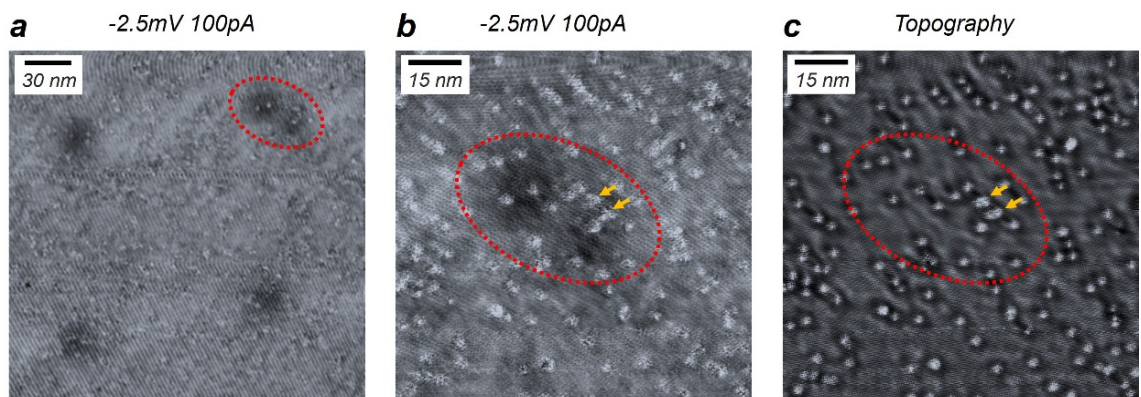

**Figure S2. Point defects near the dimer-like vortex.** a) Differential conductance map of a dimer-like vortex (red dotted ellipse) and normal vortices. b) Detailed conductance map of dime-like vortex. c) Topography image of point like defects near one of the cores of dimer-like vortex (-2.5 mV, 100pA).

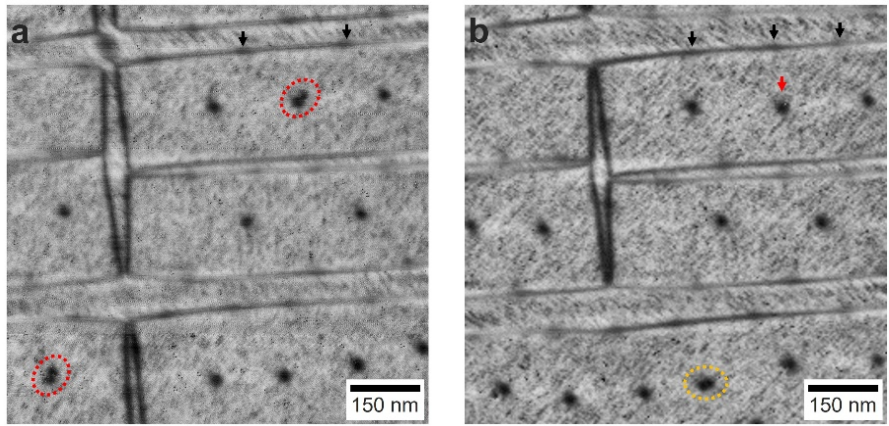

**Figure S3. Dimer-like vortex and surrounding vortices.** a) differential conductance map of dimer-like vortices (red ellipses) and normal vortices near and on twin boundaries. b) changing the dimer-like vortex to the normal vortex (red arrow) after changing the configuration of surrounding vortices near the dimer-like vortex (black arrows).

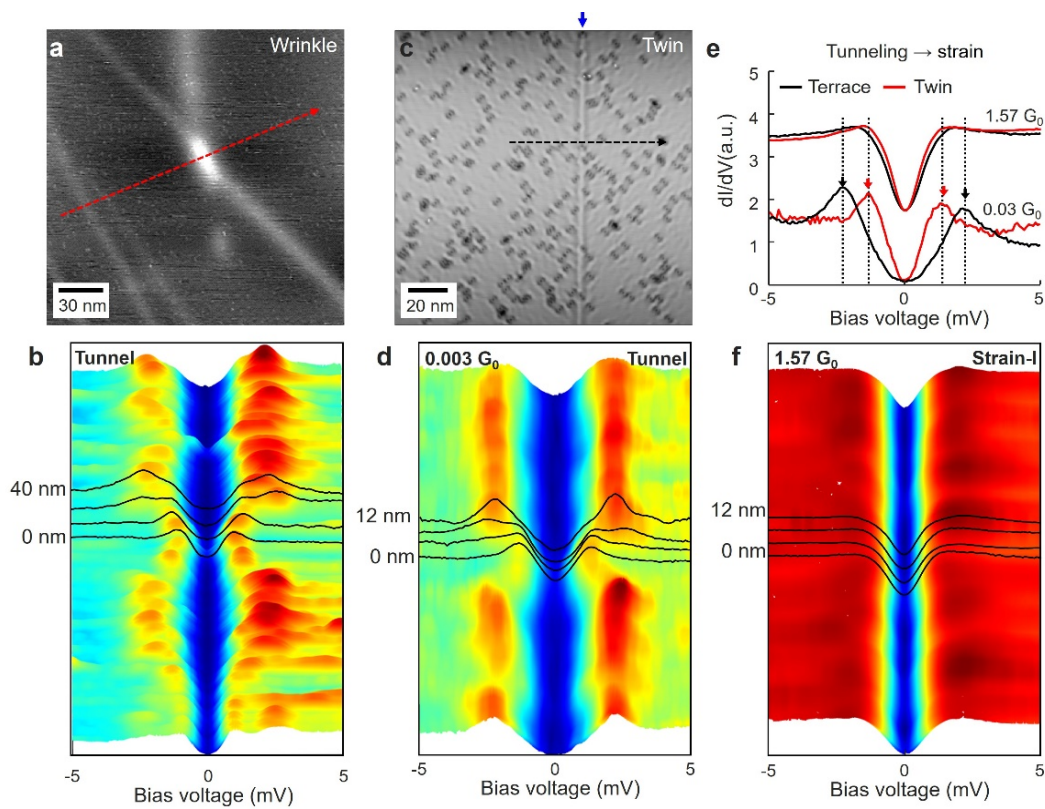

**Figure S4. Reduction of superconducting gaps near a wrinkle, a twin boundary, and at strain induced area.** a) Topographic image of the wrinkle on FeSe. b) The spatial variation of superconducting gap across the wrinkle (red dashed line in (a)). c) Topographic image of the twin boundary on FeSe (blue arrow). d) Spatial variation of superconducting gap across the twin boundary (black dashed line in (c)) in the tunneling regime. e) dI/dV spectra obtained at low conductance and high conductance on the terrace (black curves) and the twin boundary (red curves). f) Spatial variation of superconducting gap across the twin boundary (black dashed line in (c)) in the strain-I regime (1.57 $G_0$).

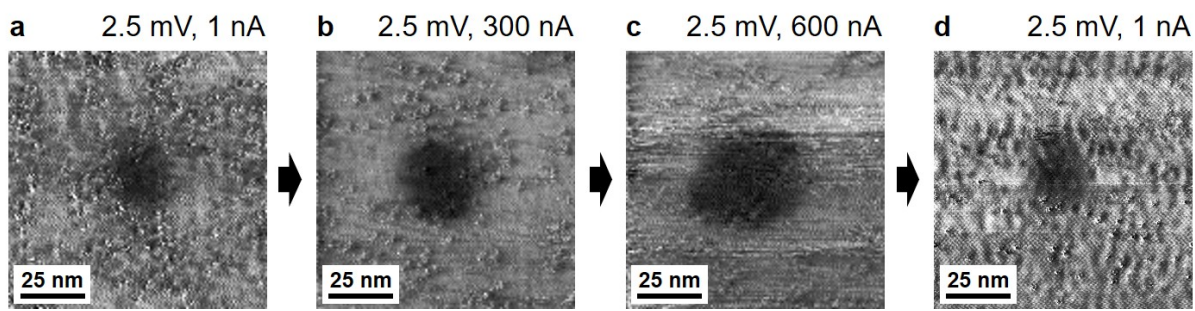

**Figure S5. Single energy conductance maps measured by sequentially changing the set point for one vortex.** These conductance maps indicate that the strain-induced modification is reversible.

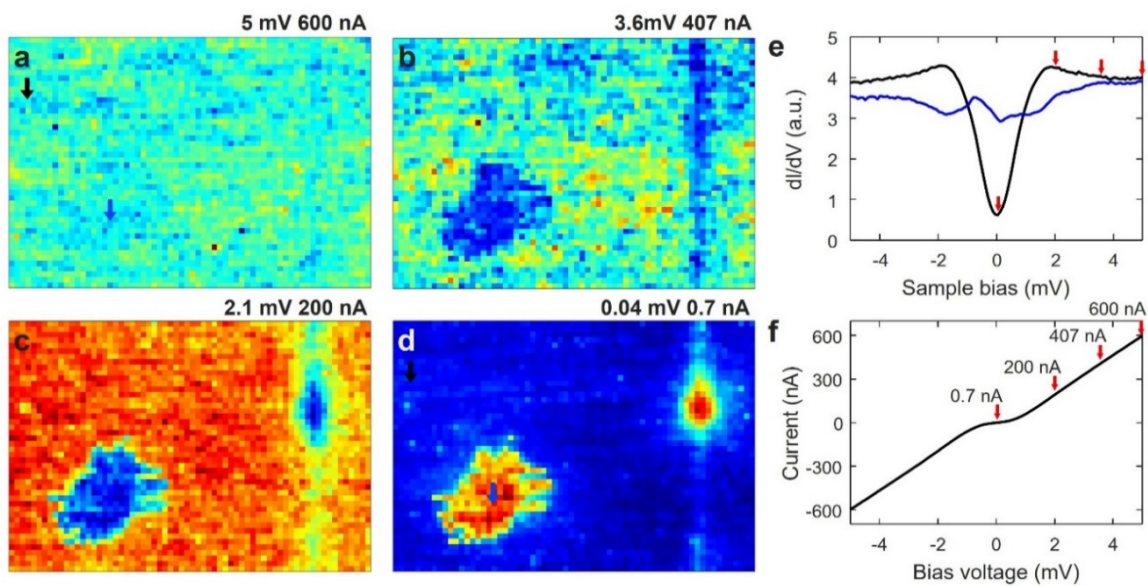

**Figure S6. The size of the expanded vortex at different biases.** This data set (main text in Figure 4d, f, and h) consists of dI/dV spectra at every point (66 x 48 points), with setting conductance at 1.57 $G_0$. During each dI/dV measurement, the current feedback loop is closed, and the tip position is fixed at the setting point (set conductance value). a-d) Differential conductance maps at 5 mV, 3.6 mV, 2.1 mV, and 0.04 mV extracted from the dI/dV mapping. e) The superconducting gaps on the terrace (black) and on the vortex core (blue). The red arrows indicate the positions of voltages at 0.04 mV, 2.1 mV, 3.6 mV, and 5mV. f) The I-V curve during measuring the dI/dV curve in (e). The red arrows indicate the bias values and corresponding currents.

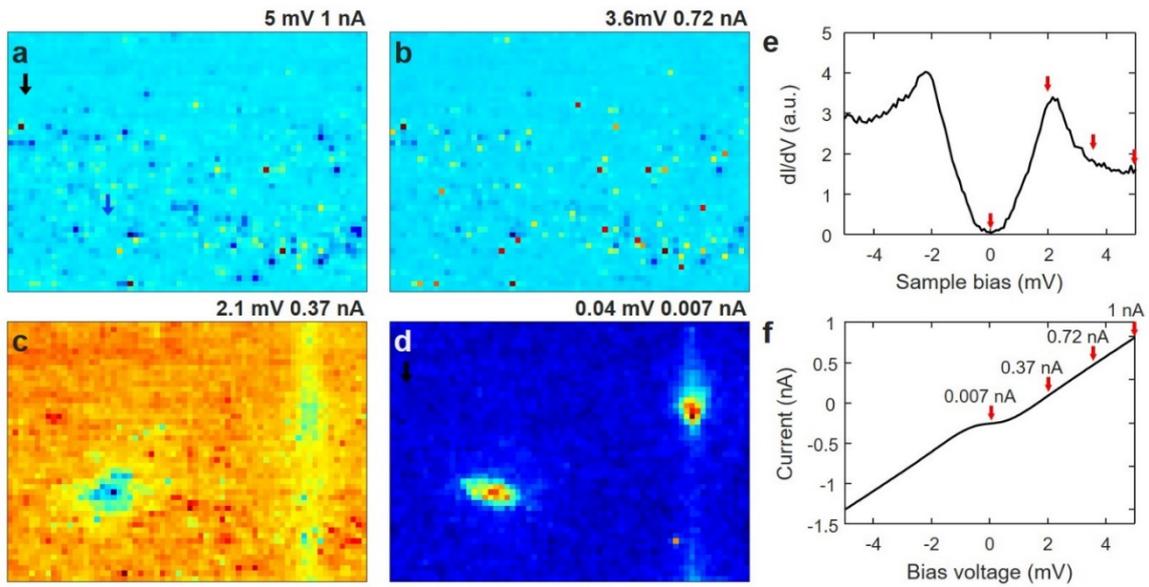

**Figure S7. The size of the normal vortex at different biases.** This data set (main text in Figure 4c, e, and g) consists of dI/dV spectra at every point (66 x 48 points), with setting conductance at 0.003 $G_0$ (tunneling). During each dI/dV measurement, the current feedback loop is closed, and the tip position is fixed at the setting point (set conductance value). a-d) Differential conductance maps at 5 mV, 3.6 mV, 2.1 mV, and 0.04 mV extracted from the dI/dV mapping. e) The superconducting gaps on the terrace. The red arrows indicate the positions of voltages at 0.04 mV, 2.1 mV, 3.6 mV, and 5mV. f) The I-V curve during measuring the dI/dV curve in (e). The red arrows indicate the bias values and corresponding currents.

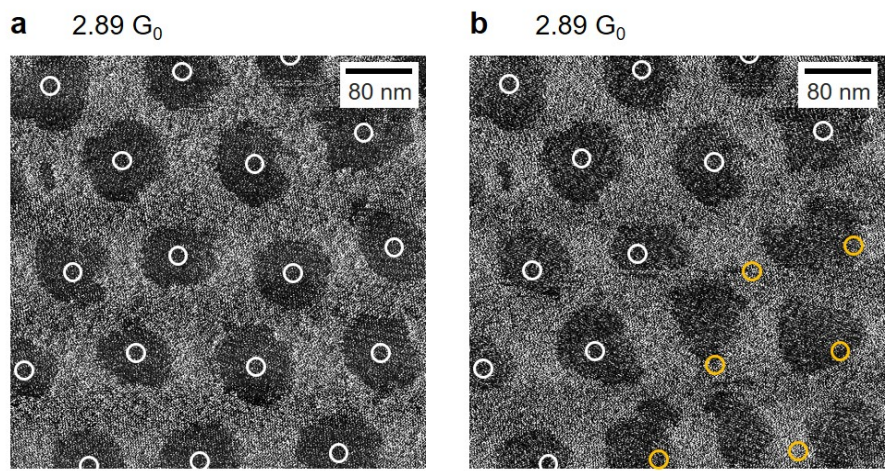

**Figure S8. Changing the partial array of vortex lattice during the high current mapping.** a) Conductance map of expanded vortices at 2.89 $G_0$ (-2.5 mV, 560 nA). b) Conductance map of locally changed vortex array (yellow dots) at 2.89 $G_0$ (-2.5 mV, 580nA).

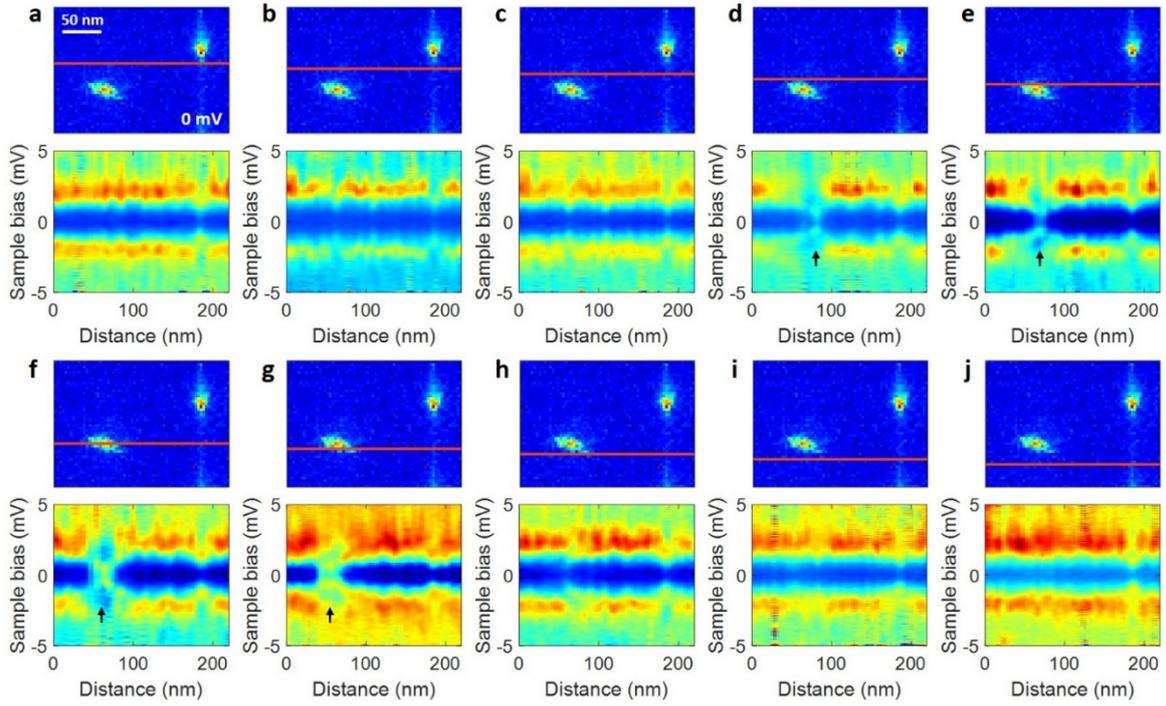

**Figure S9.** a-j) Zero bias conductance maps that contain a vortex on the terrace and a vortex on the twin boundary in tunneling regime (0.003 $G_0$), and the spatial variations of superconducting gaps across the vortex on the terrace in tunneling regime (0.003 $G_0$).

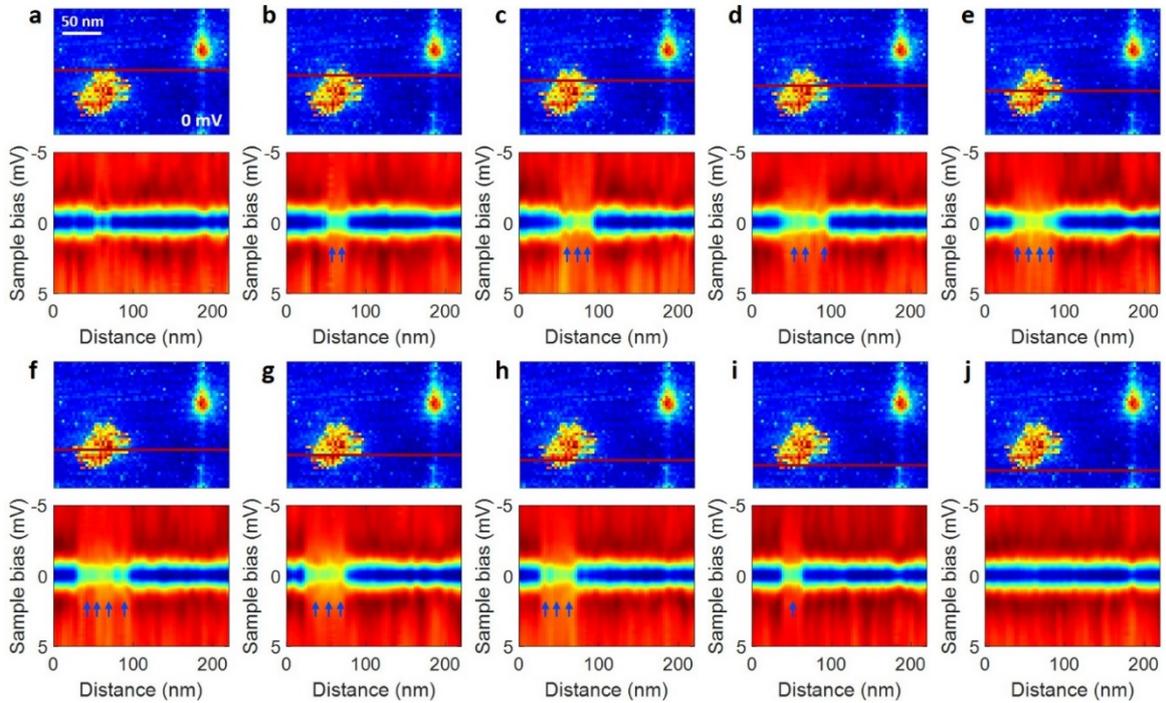

**Figure S10.** a-j) Zero bias conductance maps that contain a vortex on the terrace and a vortex on the twin boundary at 1.57 $G_0$ ($V_{set}$ = 5 mV, $I_{set}$ = 600 nA), and the spatial variations of superconducting gaps across the vortex on the terrace at 1.57 $G_0$.

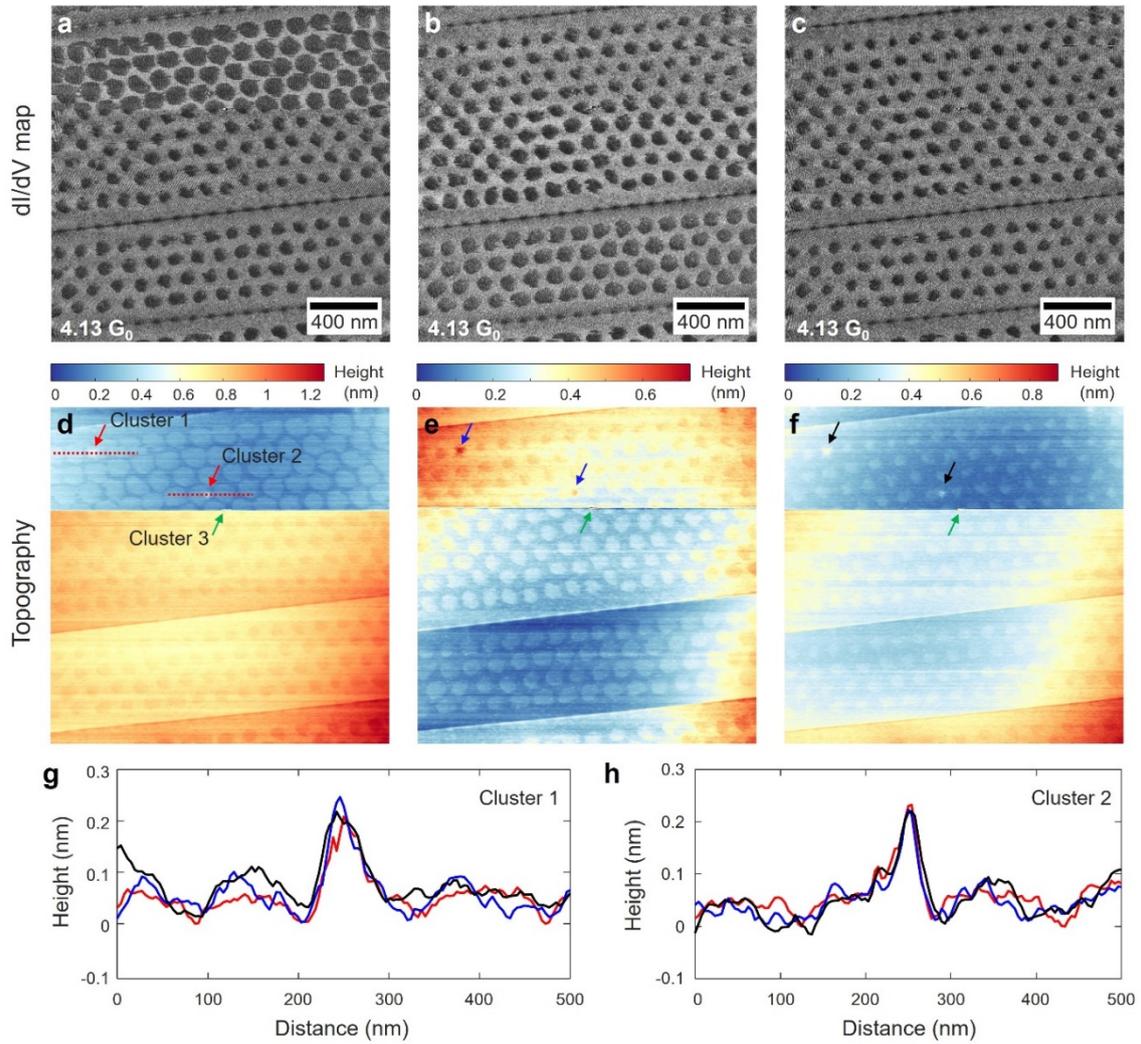

**Figure S11. Sensitivity of vortex displacement to changing STM tip structure.** a-c) Differential conductance maps of expanded vortices as tip changes at 2.5 mV, 800 nA (4.13 $G_0$). d-f) Topographic images of the expanded vortices, twin boundaries, and clusters as tip changes (setpoint: $V_{bias}$ = 2.5 mV, I = 800 nA). There are three clusters (red and green arrows in (d). The cluster 3 (green arrow in (d)) continuously changes the tip apex during the differential conductance mapping. g, h) Topographic line profiles across the cluster 1 and cluster 2 for the different tip apexes in (d), (e), and (f).

# Supplementary Note 1 – Analytical description of expanded vortex core

For the size of inclusion comparable to the coherence length ($a_x \sim \xi$), the vortex binding energy to an anisotropic inclusion is given by[1-3]

$$U_p \approx \varepsilon_0 a_z \ln\left(1 + \frac{a_x^2}{\xi_{ab}^2}\right) \quad (1)$$

Where $\varepsilon_0 = \Phi_0^2/(4\pi\lambda)^2$ is the typical vortex energy scale, $\lambda$ is the London penetration depth, the $a_x$ is the inclusion size of the x-axis, $a_z$ is the inclusion size of the z-axis, and $\xi_{ab}$ is the coherence length for the ab-plane. We can obtain the pinning force $f_p$ by dividing the vortex pinning energy by the inclusion size $a_x$.

$$f_p \approx \varepsilon_0 \frac{a_z}{a_x} \ln\left(1 + \frac{a_x^2}{\xi_{ab}^2}\right) \quad (2)$$

Furthermore, Willa, R. et al.[4] adapted dimensionless functions $G_i(\gamma a_z/a_x)$ to pinning force to describe the dependence of size and anisotropy for the large defects ($a_x \gg \xi$). The dimensionless functions contain the defect's aspect ratio ($a_z/a_x$) and the anisotropy parameter $\gamma$ ($\gamma = \xi_{ab}/\xi_c$) of the superconductor. In a similar way, we can adapt the dimensionless functions $G_i(\gamma a_z/a_x)$ to equation (2) which describes a relatively small defect ($a_x \sim \xi$).

$$f_p(a_x, a_z, \gamma) \approx G_1\left(\frac{\gamma a_z}{a_x}\right)\frac{\varepsilon_0}{\gamma} \ln\left(1 + G_2\left(\frac{\gamma a_z}{a_x}\right)\frac{a_x^2}{\xi_{ab}^2}\right) \quad (3)$$

The deformation of the vortex line is determined by the balance between the elasticity of the vortex line and the pinning force,[4,5]

$$f_p(x) = \bar{C} u \quad (4)$$

Where $\bar{C}$ is the effective spring constant and relates to the elastic Green's function.[6] For a small pinning center, $\bar{C} \approx 3\sqrt{\varepsilon_1 \varepsilon_0}/a_\Delta$ is constant.[4,5] Where, $\varepsilon_1 \approx \varepsilon_0/\gamma^2$ is the vortex line tension, and $a_\Delta$ is the vortex lattice constant (~ 140 nm at 0.13 T on FeSe). From these results, we fitted the dependence of deformations of vortex lines on the conductance values obtained from various tips (red, black, and blue circles in Figure 5c in main text) with the form of $A \cdot \ln(1 + B \cdot x^2)$ (red, black, and blue dotted lines in Figure 5c in main text and Figure S12). We confirmed that the strength of deformation of the vortex line is determined by the defect's aspect ratio ($a_z/a_x$).

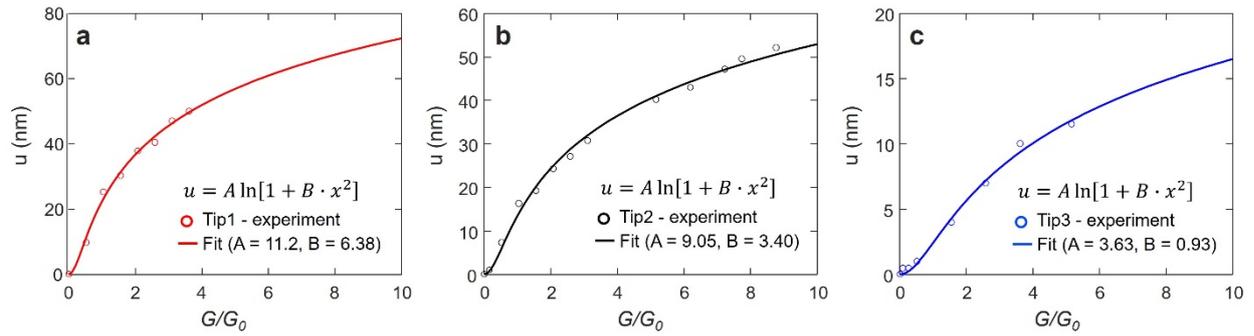

**Figure S12.** a-c) Dependence of vortex line deformation, u, on conductance values obtained from tip1 (red), tip2 (black), and tip3 (blue), with line fitting based on $A \ln[1 + B \cdot x^2]$.

References
[S1] C. J. van der Beek, M. Konczykowski, A. Abal'oshev, I. Abal'osheva, P. Gierlowski, S. J. Lewandowski, M. V. Indenbom, S. Barbanera, *Phys. Rev. B* **2002**, 66, 024523.
[S2] A. E. Koshelev, A. B. Kolton, *Phys. Rev. B* **2011**, 84, 104528.
[S3] C. J. van der Beek, M. Konczykowski, R. Prozorov, *Supercond. Sci. Technol.* **2012**, 25, 084010.
[S4] R. Willa, A. E. Koshelev, I. A. Sadovskyy, A. Glatz, *Phys. Rev. B* **2018**, 98, 054517.
[S5] R. Willa, A. E. Koshelev, I. A. Sadovskyy, A. Glatz, *Supercond. Sci. Technol.* **2018**, 31, 014001
[S6] G. Blatter, V. B. Geshkenbein, J. A. G. Koopmann, *Phys. Rev. Lett.* **2004**, 92, 067009.